\documentclass[11pt,preprint]{aastex}
\usepackage{bm}
\usepackage{epsfig}

\begin{document}

\title{\bf Large-scale peculiar motions and cosmic acceleration}

\author{Christos G. Tsagas}

\affil{Section of Astrophysics, Astronomy and Mechanics, Department of Physics,\\ Aristotle University of Thessaloniki, Thessaloniki 54124, Greece}

\date{\empty}

\begin{abstract}
Recent surveys seem to support bulk peculiar velocities well in excess of those anticipated by the standard cosmological model. In view of these results, we consider here some of the theoretical implications of large-scale drift motions. We find that observers with small, but finite, peculiar velocities have generally different expansion rates than the smooth Hubble flow. In particular, it is possible for observers with larger than the average volume expansion at their location, to experience apparently accelerated expansion when the universe is actually decelerating. Analogous results have been reported in studies of inhomogeneous (nonlinear) cosmologies and within the context of the Lemaitre-Tolman-Bondi models. Here, they are obtained within the linear regime of a perturbed, dust-dominated Friedmann-Robertson-Walker cosmology.
\end{abstract}


\section{Introduction}
In idealised Friedmann-Robertson-Walker (FRW) cosmologies, comoving observers simply follow the universal expansion. In more realistic models, the smooth Hubble flow is distorted and matter acquires `peculiar' velocities. The dipolar anisotropy of the Cosmic Microwave Background (CMB) has been traditionally interpreted as the result of our peculiar motion relative to the cosmic rest-frame: the frame that redshifts with the expansion and in which the dipole vanishes. Our Local Group of galaxies drifts with respect to the CMB frame at roughly 600~km/sec~\citep{P,D,SW}. Analogous velocities, but for bulk motions on much larger scales, were also recently reported in the surveys of~\cite{WFH,FWH} and those of~\cite{KA-BKE1,KA-BKE2,KA-BEEK}. The latter group, in particular, finds coherent peculiar flows as strong as 1000~km/sec out to scales of 450 and 800~Mpc. Both surveys appear in disagreement with the current concordance $\Lambda$CDM scenario (e.g.~see~\cite{Pe}).

This report considers the theoretical implications of such drift velocities for the kinematics of the associated observers by focusing on the scalars that describe their average volume expansion. The key question is whether observers drifting relative to the CMB and those following the Hubble expansion (in a dust-dominated FRW universe) can `measure' different deceleration parameters. Whether, in particular, it is theoretically possible for a peculiarly moving observer to `experience' accelerated expansion while the universe is decelerating. We show that, even when the peculiar velocities are relatively small, the answer to this question is positive and explain why. Not surprisingly, we also find that the effects of the peculiar motions are local. Nevertheless, the affected scales can be large enough to give the impression that the universe had recently moved into an accelerating phase. Another way of interpreting our results is that accelerated expansion for an observer moving relative to the CMB does not necessarily imply the same for the universe itself.

\section{Drift motions in perturbed FRW universes}

\begin{figure}[tbp]
\begin{center}
\includegraphics[height=2.5in,width=5.5in,angle=0]{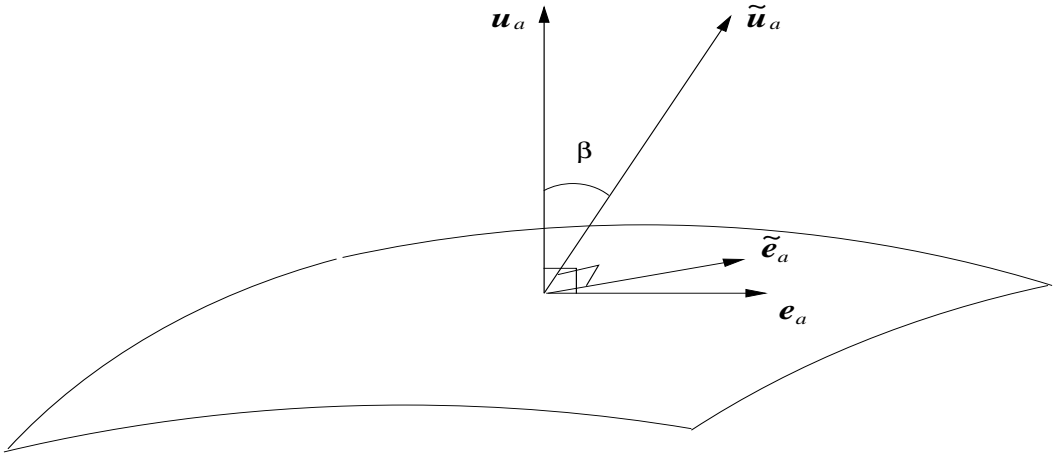}\quad
\end{center}
\caption{Observers with 4-velocity $\tilde{u}_a$ and peculiar velocity $v_a=ve_a$, relative to the reference $u_a$-frame. The (hyperbolic) angle $\beta$ determines the `tilt' between $u_a$ and $\tilde{u}_a$.}  \label{fig:mfluid}
\end{figure}

The Microwave Background introduces a preferred cosmological frame, relative to which large-scale peculiar velocities can be defined and measured. If $u_a$ is the reference 4-velocity of the CMB, typical observers in the universe have (see Fig.~\ref{fig:mfluid})
\begin{equation}
\tilde{u}_a= u_a+v_a\,,  \label{eq:tua}
\end{equation}
where $v_a$ (with $u_av^a=0$ and $v^2=v_av^a\ll1$) is their drift velocity~\citep{KE}.\footnote{The $\tilde{u}_a$-field is also timelike, since $\tilde{u}_a\tilde{u}^a=-1$ irrespective of the magnitude of the peculiar velocity. Each frame defines its own time direction and 3-space (parallel and orthogonal to the corresponding 4-velocity respectively). The tensors $h_{ab}=g_{ab}+u_au_b$ and  $\tilde{h}_{ab}=g_{ab}+ \tilde{u}_a\tilde{u}_b$, with $g_{ab}$ representing the spacetime metric, project orthogonal to $u_a$ and $\tilde{u}_a$ respectively. These tensors also define the orthogonally projected covariant derivative operators by means of ${\rm D}_a=h_a{}^b\nabla_b$ and $\tilde{\rm D}_a= \tilde{h}_a{}^b\nabla_b$ ($\nabla_a$ is the standard covariant derivative)~\citep{EvE,TCM}.} The CMB also defines the coordinate system where the universe is a dust-dominated FRW model. The `tilded' frame, on the other hand, corresponds to a typical observer in a galaxy like the Milky Way.

The average kinematics of the tilded observers are determined by the volume scalar ($\tilde{\Theta}= \nabla^a\tilde{u}_a$) of their motion~\citep{EvE,TCM}. Positive values for $\tilde{\Theta}$ imply that the mean separation between these observers increases and therefore indicate expansion. Similarly, $\Theta$ (with $\Theta=\nabla^au_a>0$) monitors the expansion of the universe. To first order in $v_a$, the two scalars are related by \citep{M}
\begin{equation}
\tilde{\Theta}= \Theta+ \vartheta\,, \label{eq:Thetas}
\end{equation}
where $\vartheta=\tilde{\rm D}^av_a$. This scalar measures the average separation between neighbouring peculiar-flow lines.\footnote{A more familiar form for Eq.~(\ref{eq:Thetas}) is the Newtonian expression $\tilde{u}_a=Hr_a+v_a$, where $\tilde{u}_a$ and $v_a$ are respectively the physical and the peculiar velocities of an observer with physical coordinates $r_a$ ($H=\Theta/3$ is the Hubble parameter). The (physical) divergence of the above leads to Eq.~(\ref{eq:Thetas}), with $\tilde{\Theta}$ and $\vartheta$ corresponding to $\partial^a\tilde{u}_a$ and $\partial^av_a$ respectively.} Expression (\ref{eq:Thetas}) implies that, in regions where $\vartheta$ is positive, the peculiarly moving observers expand faster than the universe (i.e.~$\tilde{\Theta}>\Theta$). For our purposes it is crucial that the drift motion `adds' to the background expansion and the reasons should become clear as we proceed. We will therefore always consider sections where $\vartheta$ is positive.

In multi-systems, each group of observers has its own time direction. So, in our case, time can be measured relative to the CMB frame and along that of the tilded observers. The rate of the expansion along a given time direction is determined by the associated Raychaudhuri equation~\citep{EvE,TCM}. When the universe is a dust-dominated FRW model and the drift velocities are small, the Raychaudhuri equations in the CMB and the tilded frames are~\footnote{We assume non-relativistic peculiar velocities and therefore drop terms of order $v^2$ and higher from (\ref{eq:Rays1}b) and the rest of our equations. Also, throughout this letter we use geometrised units with $c=1=8\pi G$.}
\begin{equation}
\Theta^{\prime}= -{1\over3}\,\Theta^2- {1\over2}\,\rho \hspace{15mm} {\rm and} \hspace{15mm} \dot{\tilde{\Theta}}= -{1\over3}\,\tilde{\Theta}^2- {1\over2}\,\tilde{\rho}+ \tilde{\rm D}^a\tilde{A}_a\,,  \label{eq:Rays1}
\end{equation}
respectively. Here, primes indicate time differentiation along $u_a$ and overdots are time derivatives with respect to the $\tilde{u}_a$-field. In other words, $\Theta^{\prime}=u^a\nabla_a\Theta$ and $\dot{\tilde{\Theta}}=\tilde{u}^a\nabla_a\tilde{\Theta}$, with the 4-velocity vectors related through Eq.~(\ref{eq:tua}). Also, $\rho$ and $\tilde{\rho}$ are the matter densities in the CMB and the tilded frames respectively (with $\tilde{\rho}=\rho$ to linear order in $v_a$ -- see~\cite{M}). Finally, $\tilde{A}_a$ is the 4-acceleration in the tilded frame. This vector vanishes in the CMB frame by definition (i.e.~$A_a=0$) but is nonzero in every other relatively moving coordinate system. In particular, to linear order in $v_a$, we find that $\tilde{A}_a= \dot{v}_a+(\Theta/3)v_a$~\citep{M}. The  4-acceleration term in Eq.~(\ref{eq:Rays1}b) is central to our analysis. Its presence means that expressions (\ref{eq:Rays1}a) and (\ref{eq:Rays1}b) are different, even when matter is in the form of pressureless dust and the peculiar velocities are small. In other words, observers drifting relative to the CMB have expansion rates different than that of the actual universe simply because of their relative motion. This represents a significant theoretical deviation from the conventional single-fluid studies (e.g.~see~\cite{HS}).

\section{The deceleration parameter of the drifting observer}
Expressed in terms of their volume scalars, the deceleration parameters associated with the $u_a$ and $\tilde{u}_a$ frames are
respectively given by
\begin{equation}
q= -\left(1+{3\Theta^{\prime}\over\Theta^2}\right) \hspace{15mm} {\rm and} \hspace{15mm} \tilde{q}= -\left(1+{3\dot{\tilde{\Theta}}\over\tilde{\Theta}^2}\right)\,.  \label{eq:qs1}
\end{equation}
Our main question is whether $\tilde{q}$ can take negative values while $q$ is still positive. If so, the tilded observers will experience accelerated expansion in a decelerating universe. To investigate this possibility, we first use definitions (\ref{eq:qs1}) to recast expressions (\ref{eq:Rays1}) into
\begin{equation}
(1+q)\Theta^2= \Theta^2+ {3\over2}\,\rho \hspace{15mm} {\rm and} \hspace{15mm} (1+\tilde{q})\tilde{\Theta}^2= \tilde{\Theta}^2+ {3\over2}\,\tilde{\rho}- 3\tilde{\rm D}^a\tilde{A}_a\,,  \label{eq:Rays2}
\end{equation}
respectively. These already show that $q$ and $\tilde{q}$ are generally different, but it helps to relate the two deceleration parameters directly. Recall that $\tilde{\rho}=\rho$ and $\tilde{A}_a=\dot{v}_a+(\Theta/3)v_a$ to linear order in $v_a$. Then, employing definition $\vartheta=\tilde{\rm D}^av_a$, relation (\ref{eq:Thetas}) and keeping up to $v_a$-order terms, expressions (\ref{eq:Rays2}a) and (\ref{eq:Rays2}b) combine to
\begin{equation}
(1+\tilde{q})\tilde{\Theta}^2= (1+q)\Theta^2+ \Theta\vartheta-
3\tilde{\rm D}^a\dot{v}_a\,,  \label{eq:tq1}
\end{equation}
where $\Theta$, $\vartheta>0$ always. We may also involve the volume scalar of the peculiar motion further by using the (linear in $v_a$) relation $\dot{\vartheta}=\tilde{\rm D}^a\dot{v}_a -\Theta\vartheta/3$~\citep{ET}. Then, Eq.~(\ref{eq:tq1}) leads to
\begin{equation}
1+ \tilde{q}= (1+q)
\left(1+{\vartheta\over\Theta}\right)^{-2}-
{3\dot{\vartheta}\over\Theta^2} \left(1+{\vartheta\over\Theta}\right)^{-2}\,,  \label{eq:tq2}
\end{equation}
given that $\tilde{\Theta}=\Theta+\vartheta$. The above relates the deceleration parameter in the tilded frame to that of the actual universe and it is our main result. It should now be clear that $q$ and $\tilde{q}$ are generally different. Moreover, as long as the right-hand side of (\ref{eq:tq2}) remains below unity, positive values for $q$ do not a priori guarantee the same for $\tilde{q}$. In other words, it is theoretically possible for the tilded observer to experience accelerated expansion in a decelerating universe.~\footnote{Expression (\ref{eq:tq2}) implies that two decelerating expansions can combine to give an accelerating one. Another way of showing this is by writing Eq.~(\ref{eq:Thetas}) as $\dot{\tilde{a}}/\tilde{a}= (\dot{a}/a)+(\dot{\alpha}/\alpha)$, where $\tilde{a}$, $a$ and $\alpha$ are the three scale factors (with $\dot{a},\,\dot{\alpha}>0$). Then, $\ddot{\tilde{a}}/\tilde{a}= (\ddot{a}/a)+(\ddot{\alpha}/\alpha)+ 2(\dot{a}/a)(\dot{\alpha}/\alpha)$, meaning that negative values for $\ddot{a}$ and $\ddot{\alpha}$ do not guarantee the same for $\ddot{\tilde{a}}$. Note that for simplicity we have used overdots for both time derivatives.} Putting it differently, one could say that measuring negative deceleration parameter in a frame drifting relative to the CMB (like that of our Local Group for example) does not necessarily imply an accelerating universe.

At this point it is worth noting that, according to (\ref{eq:qs1}b), condition $-1<\tilde{q}<0$ is equivalent to $-\tilde{\Theta}^2/3<\dot{\tilde{\Theta}}<0$. This means that both $\tilde{q}$ and $\dot{\tilde{\Theta}}$ can be simultaneously negative. Analogous relations also hold between $q$, $\Theta^2$ and $\Theta^{\prime}$. Given that, one should distinguish between accelerated expansion with simply $-1<\tilde{q}<0$ and that with $\dot{\tilde{\Theta}}>0$. We may therefore view $-1<\tilde{q}<0$ and $\dot{\tilde{\Theta}}>0$ (equivalently $\tilde{q}<-1$) as the conditions for `weakly'  and `strongly' accelerated expansion respectively. Then, it is important to recognise that, as long as we only require $\tilde{q}$ to lie in the (-1,0) range, the 4-acceleration term in Eqs.~(\ref{eq:Rays1}b) and (\ref{eq:Rays2}b) does not need to dominate the right-hand side of these expressions. This implies that peculiar motions can lead to weakly accelerated expansion within the limits of the linear (the almost-FRW) approximation. Given that, we will focus on the $-1<\tilde{q}<0$ case for the rest of this letter. Note that the supernovae results put the deceleration parameter close to $-0.5$~\citep{TR,Retal2}.

\begin{figure}[tbp]
\begin{center}
\includegraphics[height=2.5in,width=7.5in,angle=0]{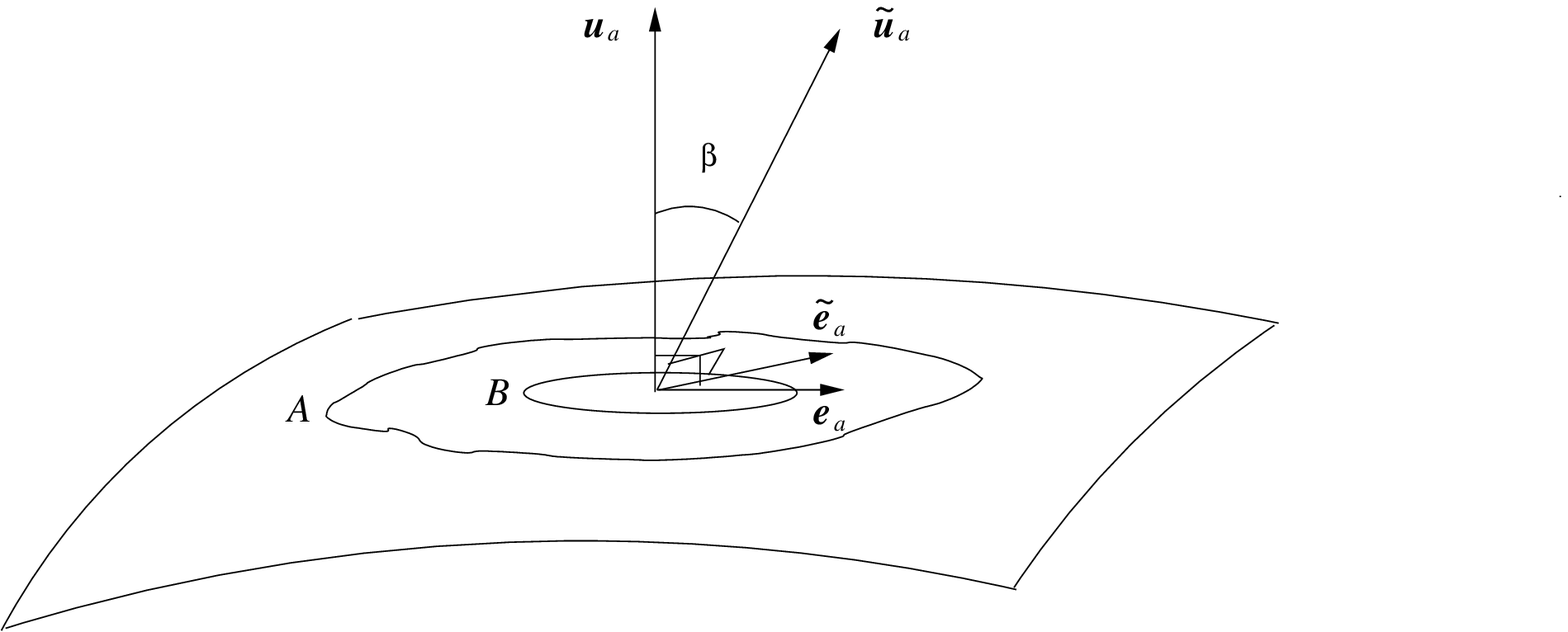}\quad
\end{center}
\caption{The patch ($A$) has positive $\vartheta=\tilde{\rm D}^av_a$ and so expands faster than its surroundings (see Eq.~(\ref{eq:Thetas})). Inside region ($B$) the right-hand side of expression (\ref{eq:tq2}) drops below unity and there the comoving observer `measures' negative deceleration parameter.}  \label{fig:pvel}
\end{figure}

\section{Apparent acceleration in perturbed FRW universes}
Let us now consider an extended spatial region ($A$) -- see Fig.~\ref{fig:pvel}, which largely complies with the FRW symmetries and expands with the Hubble flow, but is still endowed with a bulk peculiar velocity field that `adds' to the background expansion (i.e.~with $\vartheta>0$). Typical observers inside ($A$) have peculiar velocities close to the mean bulk flow of the patch. To linear order in $v_a$, the deceleration parameter for those observers obeys Eq.~(\ref{eq:tq2}). The simplest case corresponds to $3\dot{\vartheta}/\tilde{\Theta}^2\simeq0$, which occurs when $\vartheta$ varies very slowly with time (for example). Then, when the Hubble expansion dominates the kinematics, $\vartheta/\Theta\ll1$ and a straightforward Taylor expansion reduces Eq.~(\ref{eq:tq2}) to
\begin{equation}
1+ \tilde{q}= \left(1+{1\over2}\,\Omega\right)
\left[1-2\left({\vartheta\over\Theta}\right)\right]\,.
\label{eq:ltq1}
\end{equation}
Recall that $q=\Omega/2$ in dust-dominated FRW models, with $\Omega=3\rho/\Theta^2$ and $\rho$ representing the density of the matter in the $u_a$-frame. Noting that $\Omega$ may also be seen as the effective density parameter of patch ($A$), the tilded observers will experience accelerated expansion if
\begin{equation}
\left(1+{1\over2}\,\Omega\right)
\left[1-2\left({\vartheta\over\Theta}\right)\right]< 1\,.  \label{eq:con1}
\end{equation}
Whether this condition is satisfied or not and the affected scale (i.e.~the size of patch ($B$) in Fig.~\ref{fig:pvel}), depends on the value of $\Omega$ and on the ratio $\vartheta/\Theta$.  To estimate the latter we need to know the bulk velocities of drift motions on scales far beyond that of our Local Group.

Peculiar velocities are difficult to measure, since direct measurements only provide their radial component. One also needs to subtract the Hubble expansion, which requires independent knowledge of the galaxy's distance. As a result, bulk peculiar velocities are estimated by means of statistical methods~\citep{SW}. Recent independent reports have claimed large-scale coherent drift velocities significantly higher than those anticipated by the concordance $\Lambda$CDM model. These surveys extend to lengths of 100$h^{-1}$Mpc~\citep{WFH,FWH}, 300$h^{-1}$~Mpc and 500$h^{-1}$~Mpc~\citep{KA-BKE1,KA-BKE2,KA-BEEK}, with $h$ being the Hubble parameter in units of 100~km/sec\,Mpc. The results show bulk velocities as large as 500~km/sec~\citep{WFH,FWH} and up to 1000~km/sec~\citep{KA-BKE1,KA-BKE2,KA-BEEK} on the corresponding scales. On smaller lengths (between 30 and 60~Mpc) the work of~\cite{LS} suggests a (positive) variance in the local Hubble rate up to 10\%. With the possible exception of the last survey, there is currently no way of knowing whether the reported bulk flows are of the desired type (i.e.~with $\vartheta>0$). Nevertheless, in the absence of better data, we will use the magnitudes of the aforementioned peculiar velocities to infer reasonable (order-of-magnitude) estimates for $\vartheta$. In addition, mainly for algebraic simplicity and illustration purposes, we will also consider the intermediate value of 700~km/sec as a yardstick peculiar velocity. Note that this value is very close to the drift velocity of our Local Group.

Setting the Hubble parameter at 70~km/sec\,Mpc and extrapolating to 50~Mpc, 100~Mpc and 1000~Mpc, we find that $\vartheta/\Theta$ is close to 1/5, 1/10 and 1/100 respectively.\footnote{Recall that $\Theta=3H$ and that $\vartheta=\tilde{\rm D}_av^a\simeq\partial_av^a$. Measuring the 3-divergence of the peculiar velocity is not feasible at present. We can obtain an estimate, however, using the approximate relation $\partial_av^a=\partial v^a/\partial r^a\sim3v/r$, where $v$ is the magnitude of the bulk velocity in a given region and $r$ the size of that region. Then, $\vartheta/\Theta\sim v/Hr$.} Then, following condition (\ref{eq:con1}), the tilded observer will `measure' negative deceleration parameter within a region of up to 50~Mpc (in an otherwise decelerating universe) if $0<\Omega<1.3$. This condition strengthens to $0<\Omega<0.5$ at 100~Mpc, while further out, near the 1000~Mpc mark for instance, $\tilde{q}$ will remain positive unless $0<\Omega<0.04$. Inserting these numbers into Eq.~(\ref{eq:ltq1}) we obtain a range of values for the deceleration parameter of the tilded observer on the corresponding scales. Thus, provided (\ref{eq:con1}) is satisfied, $\tilde{q}$ varies within (-0.4,\,0) on scales of 50~Mpc, between (-0.2,\,0) when we move to 100~Mpc and within (-0.02,\,0) near the 1000~Mpc threshold. So, in this example the size of accelerated region (i.e.~that of patch ($B$) in Fig.~\ref{fig:pvel}) ranges between 50~Mpc and 1000~Mpc. Within these scales $\tilde{q}$ lies in the (-0.4,\,0) range, taking its minimum value in small-scale regions of low density and approaching zero as we move on to larger lengths. These estimates are not far from those inferred by the supernovae data, which value the deceleration parameter close to -0.5 and put the transition to deceleration near $z=0.5$~\citep{TR,Retal2}. The picture does not change much when we adopt the results of~\cite{LS}, the surveys of~\cite{WFH} and~\cite{FWH}, or those of~\cite{KA-BKE1,KA-BKE2,KA-BEEK}. Substituted into expressions (\ref{eq:ltq1}) and (\ref{eq:con1}), the former give $-0.2<\tilde{q}<0$ in regions of 50~Mpc when $\Omega<0.5$ there. Similarly, close to 150~Mpc, the measurements of~\cite{WFH} and~\cite{FWH} put $\tilde{q}$ in the range (-0.1,\,0), provided $\Omega<0.2$ there. Finally on lengths of 450 and 800~Mpc, the results of~\cite{KA-BKE1,KA-BKE2,KA-BEEK} suggest that $\tilde{q}$ varies the range (-0.07,\,0) and (-0.04,\,0) respectively, when $\Omega<0.15$ and $\Omega<0.07$ on the corresponding scales. Note that the same survey indicates bulk flows of 1500~km/sec on scales close to 150~Mpc. Inserted into Eqs.~(\ref{eq:ltq1}), (\ref{eq:con1}) these values lead to $-0.3<\tilde{q}<0$ when $\Omega<0.8$. One should keep in mind, however, that on relatively small scales the peculiar-velocity errorbars are large (see~\cite{KA-BKE1}).

Let us now turn to the last term of Eq.~(\ref{eq:tq2}). Qualitatively speaking, a positive $\dot{\vartheta}$ will assist the acceleration, relax the above given conditions and lead to lower values of $\tilde{q}$. So, here, we will assume that $\dot{\vartheta}$ is negative. We will also demand that $\dot{\vartheta}/\Theta^{\prime}\simeq\vartheta/\Theta\ll1$, to ensure that both $\vartheta$ and $\dot{\vartheta}$ are small perturbations relative to their background associates. The next step is to recast Raychaudhuri's formula (see Eq.~(\ref{eq:Rays1}a)) in the form
\begin{equation}
\Theta^{\prime}=-{1\over3}\,\Theta^2
\left(1+{1\over2}\,\Omega\right)\,,  \label{eq:Ray}
\end{equation}
with $\Theta^{\prime}<0$. Solving the above for $\Theta^2$, substituting into Eq.~(\ref{eq:tq2}) and employing some straightforward algebra we arrive at
\begin{equation}
1+ \tilde{q}= \left(1+{1\over2}\,\Omega\right)
\left(1-{\vartheta\over\Theta}\right)\,.  \label{eq:ltq2}
\end{equation}
Using the previous values of $\vartheta/\Theta$, we find that negative $\tilde{q}$s on $\sim50$~Mpc scales need $\Omega<0.5$. Similarly, expression (\ref{eq:ltq2}) translates into $\Omega<0.2$ close to 100~Mpc and into $\Omega<0.02$ near the 1000~Mpc mark, if $\tilde{q}$ is to become negative there. Under these conditions, the accelerated patch extends from 50~Mpc to 1000~Mpc, with $\tilde{q}$ varying within (-0.2,\,0). So, even with the last term of (\ref{eq:tq2}) accounted for (and in an unfavourable way), negative values for $\tilde{q}$ are still possible. Conventional almost-FRW kinematics can accommodate accelerated expansion.

\section{Summary and discussion}
To summarise, suppose that in a dust-dominated FRW universe a sufficiently large region ($A$) is endowed with a weak bulk peculiar velocity of positive divergence (i.e.~$\vartheta>0$). When the right-hand side of (\ref{eq:tq2}) drops below unity, around every point in ($A$) there is an essentially spherically symmetric patch ($B$) where the expansion is `weakly' accelerated (i.e.~$-1<\tilde{q}<0$ there).\footnote{This conclusion has been based on the average peculiar kinematics without incorporating anisotropies. For instance, the symmetry of region ($A$) and the observers position in it can induce anisotropy in the spatial distribution of $\tilde{q}$. Generally, the higher the spherical symmetry of ($A$) and the closer the observer at the centre the better. These matters are less of an issue, however, when ($A$) is considerably larger than ($B$), namely as long as patch ($B$) lies well within region ($A$). The direction of the peculiar motion can also introduce an anisotropy in the $\tilde{q}$-distribution. This effect is maximised when the peculiar velocity maintains the same magnitude and direction throughout the integration period (i.e.~from $z\simeq0.5$ to the present -- see expression (\ref{eq:DL})). In the opposite case, when the $v_a$-field has been sufficiently randomised, the anisotropy will be negligible. Estimating effects like these is currently impossible, however, as it requires detailed data on the distribution of peculiar velocities within regions of several hundred Mpc.} As a result, nearly every observer in ($A$) will experience accelerated expansion, although region ($A$) and the universe itself may be actually decelerating. The accelerating effect, in a given region, depends on the magnitude of the peculiar velocity and the density of the region in question. Overall, the larger the drift velocity and the lower the density the faster the acceleration.

Little more than a decade ago, observations of high-redshift supernovae indicated that our universe was expanding at an accelerating pace~\citep{Retal1,Petal}. This conclusion was reached after applying the observed luminosity distances of the supernovae to the distance-redshift relation,
\begin{equation}
D_L= (1+z)H_0^{-1} \int_0^ze^{-\int_0^x(1+q)\mathrm{d}[\ln(1+y)]}\mathrm{d}x\,,  \label{eq:DL}
\end{equation}
of an FRW model. Note that in the above $q$ is the deceleration parameter of the universe and not that of an observer moving relative to the CMB.\footnote{To account for the effects of our peculiar motion on the right-hand side of Eq.~(\ref{eq:DL}), one should replace $q$ with $\tilde{q}$. To linear order in $v_a$, the latter is given by expression (\ref{eq:tq2}), or by its simplified counterpart (\ref{eq:ltq1}).} The results have repeatedly given negative values to $q$, indicating an accelerated expansion for our universe. In particular, the deceleration parameter was estimated close to -0.5. The same measurements also suggested that the accelerated phase was a relatively recent event, putting the transition from deceleration to acceleration around $z=0.5$ (i.e.~between two and three thousand Mpc -- \cite{TR,Retal2}). The supernovae results were so unexpected that they have since dominated almost every aspect of contemporary cosmology. The main problem is that negative values for the deceleration parameter appear theoretically  impossible in FRW (as well as in perturbed, almost-FRW) cosmologies, unless new physics or drastic changes to the matter content of the universe were introduced. Dark energy, an unknown and elusive form of matter with negative gravitational mass, has so far been the most popular answer.

The implications of peculiar velocity perturbations on the luminosity distance of distant galaxies, within the context of perturbed FRW models, has been investigated in the past (e.g.~\cite{VFW}), in an attempt to reconcile expression (\ref{eq:DL}) with positive values for the deceleration parameter. That work has investigated the impact of peculiar motions on $D_L$. Here, we have followed a different approach. Turning our attention to the deceleration parameter, the aim was to examine whether peculiar motions can `make' the latter negative. Our results show that this is theoretically possible. Peculiar motions can locally mimic the kinematic effects of dark energy. Observers moving relative to the smooth Hubble flow can have local expansion rates appreciably different than that of the actual universe. This reflects the fact that the Raychaudhuri equations in the two coordinate systems (that of the CMB and that of a drifting observer) are not the same. The difference is due to a 4-acceleration term, which vanishes in the CMB frame but takes nonzero values in any other relatively moving reference system. As a result, accelerated expansion is possible even when the drift velocities are small and matter is simple pressure-free dust, namely within the limits of the linear (almost-FRW) approximation.

Extrapolating our drift velocity relative to the CMB frame, we found that peculiarly moving observers can measure negative deceleration parameter on scales between (roughly) 50 and 1000~Mpc, with $\tilde{q}$ varying in the range (-0.4,\,0). Based on the surveys of~\cite{WFH},~\cite{FWH} and particularly those of~\cite{KA-BKE1,KA-BKE2,KA-BEEK}, the deceleration parameter was confined within (-0.3,\,0). These results are qualitatively in the right direction, though quantitatively stop short of fully reproducing the current supernovae data. In particular, the largest scale considered here is between a half and a third of the typical scale of the accelerated domain~\citep{TR,Retal2}. Also, typically, the larger the scale the lower the required (effective) density parameter, putting the latter potentially at odds with the current observational constraints. On these grounds, this work should be seen as a proof of principle, rather than a full fit to the current supernovae data. Nonetheless, it is important to know that (based on estimates inferred from peculiar-flow observations) apparently accelerating expansion is possible in linearly perturbed FRW models with conventional (dust) matter.

Given that peculiar velocities are a byproduct of structure formation, their role can be seen as a `backreaction' effect (e.g.~see~\cite{R1,R2,BMR,CPZ,KMR,W4} and also~\cite{B2} for a recent review). These scenarios consider the overall impact of inhomogeneity and anisotropy, go beyond the linear regime and usually employ an averaging scheme~\citep{B1,Z}. Averaging also raises issues related to the `fitting problem' and the choice of a `background'~\citep{E1,KMM}, the `dressing' of cosmological parameters~\citep{BC1,BC2}, the propagation of light~\citep{R3,R4} and the `synchronisation of clocks'~\citep{W1,W2,W3}. Here, we have focused on peculiar motions without introducing any spatial averaging. We have also remained within the linear approximation. Nevertheless, the effects are of the same nature. Also, since we have looked at peculiar velocities that increase the volume expansion at the observer's location, our model shares close analogies with a local void. The possibility of apparent acceleration in the void scenario has been studied by many authors both qualitatively and quantitatively (e.g.~see~\cite{MHE},%
~\cite{C},~\cite{T1,T2},~\cite{INN},~\cite{AAG1,AAG2},~\cite{ZMS},%
~\cite{CFL},~\cite{BA},~\cite{S} for a representative though incomplete list). Whereas the effects of large voids have been generally studied in the context of the Lemaitre-Tolman-Bondi (LTB) solution,\footnote{The reader is directed to~\cite{PK} for a general discussion on LTB cosmologies and also to~\cite{Zi} and~\cite{CCF} for perturbative studies of these models.} our analysis has been performed within a perturbed FRW model. The analogies between the two approaches found here, also seem to support the claim by~\cite{EMR}: that LTB models fitting the supernovae data (with appropriate initial conditions) are equivalent to perturbed FRW spacetimes along the observer's past light-cone. Although single void models appear unrealistic, given the complexity of the observed structure, the simple analysis and the results presented here suggest that (when more realistic averages are performed) identifying the deceleration parameter measured in the frame of a drifting observer with that of the universe itself could be misleading.

\noindent{\bf Acknowledgements} The author wishes to thank George Ellis for helpful discussions.

\end{document}